\documentstyle[amssymb,epsfig,prl,aps]{revtex}
\begin{document}

%TCIMACRO{
%\TeXButton{TeX field}{\articletitle{Probing interactions in~  \\ mesoscopic gold wires}
%\author{F.\ Pierre, H. Pothier, D. Esteve, and M. H. Devoret}
%\affil{Service de Physique de l'Etat Condense,
%Commissariat a l'Energie Atomique, Saclay,
%F-91191 Gif-sur-Yvette, France}
%\author{A.\ B.\ Gougam
%and Norman O. Birge}
%\affil{Department of Physics and Astronomy, 
%Michigan State University, East Lansing, MI\ 48824-1116, USA}
%\begin{abstract}
%We have measured in gold wires the energy exchange rate between
%quasiparticles, the phase coherence time of quasiparticles and the
%resistance vs. temperature, in order to probe the interaction processes which are relevant
%at low temperatures. We find that the energy exchange rate is higher than
%expected from the theory of electron-electron interactions, and that it has
%a different energy dependence. The dephasing time is constant at
%temperatures between 8 K and 0.5 K, and it increases below 0.5 K.\ The
%magnetoresistance is negative at large field scales, and the resistance
%decreases logarithmically with increasing temperatures, indicating the
%presence of magnetic impurities, probably Fe. Whereas resistivity and phase
%coherence measurements can be attributed to magnetic impurities, the
%question is raised whether these magnetic impurities could also
%mediate energy exchanges between quasiparticles.
%\end{abstract}}}%
%BeginExpansion
%\articletitle{Probing interactions in~  \\ mesoscopic gold wires}
\title{Probing interactions in mesoscopic gold wires}
\author{F.\ Pierre, H. Pothier, D. Esteve, and M. H. Devoret}
%\affil{Service de Physique de l'Etat Condens\'e,
\address{Service de Physique de l'Etat Condens\'e,
Commissariat \`a l'Energie Atomique, \\ Saclay,
F-91191 Gif-sur-Yvette, France}
\author{A.\ B.\ Gougam
and Norman O. Birge}
%\affil{Department of Physics and Astronomy, 
\address{Department of Physics and Astronomy, 
Michigan State University,\\ East Lansing, MI\ 48824-1116, USA}
\maketitle
\begin{abstract}
We have measured in gold wires the energy exchange rate between
quasiparticles, the phase coherence time of quasiparticles and the
resistance vs. temperature, in order to probe the interaction processes which are relevant
at low temperatures. We find that the energy exchange rate is higher than
expected from the theory of electron-electron interactions, and that it has
a different energy dependence. The dephasing time is constant at
temperatures between 8 K and 0.5 K, and it increases below 0.5 K.\ The
magnetoresistance is negative at large field scales, and the resistance
decreases logarithmically with increasing temperatures, indicating the
presence of magnetic impurities, probably Fe. Whereas resistivity and phase
coherence measurements can be attributed to magnetic impurities, the
question is raised whether these magnetic impurities could also
mediate energy exchanges between quasiparticles.
\end{abstract}%
%EndExpansion

Several recent experiments have demonstrated that the low-energy properties
of quasiparticles in metallic thin films are sample-dependent. On the one
hand, the power-law increase of the phase-coherence time with decreasing temperature, predicted by the
theory of electron-electron interactions in diffusive wires \cite{AA}, has
been observed in several experiments \cite{Echternach,WLGoteborg}. The
energy exchange rates between electrons was found to be in agreement with
this theory in experiments on silver wires \cite{relaxAg}. On the other hand,
the dephasing rate of quasiparticles was found to saturate at low
temperature in a series of gold wires \cite{MW}, and the energy exchange
rate between quasiparticles in copper wires to display an energy dependence
different from the predicted one \cite{relax}. We present here measurements on
gold samples, in which the energy exchange rates have the same energy
dependence as was observed in copper and have a magnitude even higher. In
order to investigate the origin of this effect, we have performed resistance
measurements on samples fabricated similarly.\ The logarithmic dependence of
the resistance, the negative magnetoresistance at large field and the
temperature dependence of the phase coherence time, which is constant
between $8~\mathrm{K}$ and $0.5~\mathrm{K}$ and increases at lower
temperature, suggest the presence of magnetic impurities, which might
mediate electron-electron interactions.

\section{Energy exchange rates}

\subsection{Measurement set-up}

\begin{figure}[!ftbh]
\centering
\epsfig{file=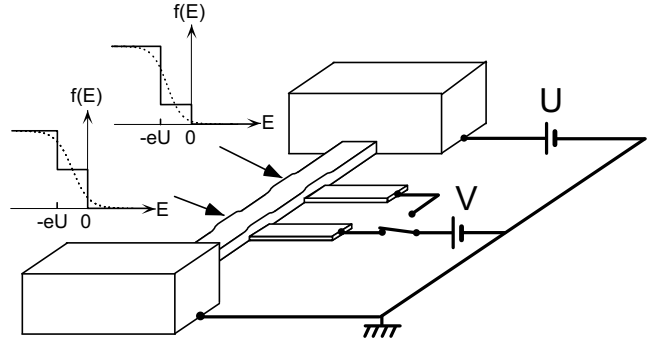, width=3.3615in, clip=}
\caption{Experimental layout: a metallic
wire of length $L$ is connected to large reservoir electrodes, biased at
potentials $0$ and $U$. In absence of interaction, the distribution function
at a distance $X=xL$ from the grounded electrode has an intermediate step $%
f\left( E\right) =1-x$ for energies between $-eU$ and $0$ (solid curves) (we
assume $U>0$). When interactions are strong enough to thermalize electrons,
the distribution function is a Fermi function (dotted curves). In the
experiment, the distribution function is obtained from the differential
conductance $dI/dV(V)$ of the tunnel junction formed by the wire and a
superconducting electrode placed underneath.}
\label{FigSetup}
\end{figure}

In order to access the energy exchange rates among quasiparticles, we have
measured the distribution function $f(E)$ in a stationary out-of-equilibrium
set-up \cite{relax}, described in Fig. \ref{FigSetup}. We consider a
mesoscopic metallic wire placed between two ideal reservoirs. In the absence
of interactions, the population of quasiparticles at a given energy
interpolates linearly between the distribution functions in the contacts,
leading, if $k_{B}T\ll eU,$ to a double-step-shaped distribution function. 
In the opposite ``hot
electron'' regime, in which the typical interaction time is much shorter
than the diffusion time $\tau _{D}=L^{2}/D,$ equilibrium is reached locally
at each position along the wire: the energy distribution function $f(x,E)$
is a Fermi function  (see dotted curves
in Fig.$~$\ref{FigSetup}). If heat is only carried out by electrons, the local
temperature is $T_{\mathrm{eff}}(x)=\sqrt{T^{2}+x\left( 1-x\right) U^{2}/%
\mathfrak{L}},$ where $\mathfrak{L}=\frac{\pi ^{2}}{3}\left( \frac{k_{B}}{e}%
\right) ^{2}$is the Lorenz number \cite{Steinbach,Kozub}. Our experiments focus on the intermediate regime,
in which interactions lead to a significant redistribution of the energy
among quasiparticles, but not to a complete thermalization. The distribution
function is obtained from the differential conductance $dI/dV\left( V\right) 
$ of a tunnel junction between the wire and superconducting electrodes \cite%
{relax,Rowell}. When the temperature of the superconductor lies well below
the superconducting transition temperature, and if the density of states in
the normal electrode wire is taken as energy independent on the probed
energy range, the differential conductance of the junction is simply
proportional to the convolution product of the density of states in the
superconductor $n_{S}(E)=\left| E\right| /\sqrt{E^{2}-\Delta ^{2}},$ with $%
\Delta $ the gap of the superconductor, and of the energy derivative of the
distribution function in the wire, $\frac{\partial f(x,E)}{\partial E}:$%
\begin{equation}
\frac{dI}{dV}\left( V\right) =\frac{-1}{R_{t}}\int n_{S}(eV-E)\frac{\partial
f(x,E)}{\partial E}dE, \label{dIdV}
%f(x,E)}{\partial E}dE,  \tag{1}  \label{dIdV}
\end{equation}%
with $R_{t}$ the tunnel resistance of the junction.

\subsection{Samples}

The samples were fabricated by deposition with an electron-gun, at several
angles, through a PMMA suspended mask patterned using e-beam lithography.
The substrate was thermally oxidized silicon, as in our experiments on Cu
and Ag. We fabricated on the same chip two Au wires: wire \#1, with length $%
L=1.55~\mu \mathrm{m,}$ and a single probe junction placed at $x=0.7$; and wire
\#2, with length $L=5~\mu \mathrm{m,}$ and two probe junctions, at $x=0.25$ and $%
x=0.5.\;$

We first deposited a $25~\mathrm{nm}$-thick aluminum film, which was then
oxidized. This layer defines the superconducting probe electrodes. The wires
and the pads were obtained by the subsequent evaporation from a 99.99\%
purity gold target, at a pressure of $10^{-6}~\mathrm{mb}$, at $1~\mathrm{%
nm/s}$. The thickness and width of the wires are $45~\mathrm{nm}$ and $165$~$%
\mathrm{nm}$. The electrodes at the ends of the wires are $500~\mathrm{nm}$%
-thick pads, with an area of about $1~\mathrm{mm}^{2}$. From the low
temperature wire resistances $R_{1}=5.39~\Omega $ and $R_{2}=16.9~\Omega ,$
we deduce from Einstein's relation, assuming rectangular cross-sections, the
diffusion constant $D=0.013~\mathrm{m}^{2}/\mathrm{s}$ and the diffusion
times $\tau _{D1}=0.18~\mathrm{ns}$ and $\tau _{D2}=1.8~\mathrm{ns.}$ The
samples were mounted in a copper box thermally anchored to the mixing
chamber of a dilution refrigerator. Electrical connections were made through
filtered coaxial lines \cite{filtres}. 

\begin{figure}[!ftbh]
\centering
\epsfig{file=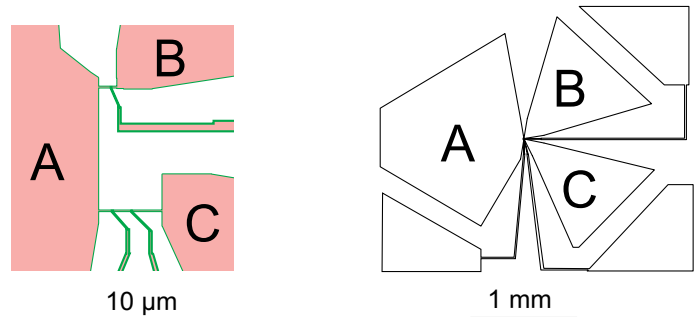, width=3.627in, clip=}
\caption{Micrometric scale (left) and large scale (right) shape of the sample.
The wires are placed between pairs of large contact pads (wire \#1 between A
and B, wire \#2 between A and C). Superconducting probe junctions are placed
on wire \#1 at $x=0.7$, on wire \#2 at $x=0.25$ and $x=0.5$. The shape of
the pads A, B, C has been designed to optimize the cooling of the reservoirs.}
\label{dessin}
\end{figure}

Measurements proceed as follows. In a first step, the voltage $U$ is set to
zero. From the comparison of the measured differential conductance of the
tunnel junctions with the calculated convolution product of the BCS density
of states and of the derivative of a Fermi function at the temperature of
the mixing chamber (see Eq.~(1)), we deduce the gap of the superconductor ($%
\Delta \approx 0.2~\mathrm{mV}$) and the tunnel resistances \cite{field} ($%
R_{t}=57~\mathrm{k\Omega }$ (wire \#1), $28~\mathrm{k\Omega }$ and $30~%
\mathrm{k\Omega }$ (wire \#2)). In a second step, the voltage $U$ is set to
a dc value. From the differential conductance of the tunnel junctions, we
deduce by numerical deconvolution the distribution functions in the wire at
the position of the junctions, using Eq.~(1) and the values of the gap and
tunnel resistance determined in the first step.

\subsection{Distribution functions}

\begin{figure}[!ftbh]
\centering
\epsfig{file=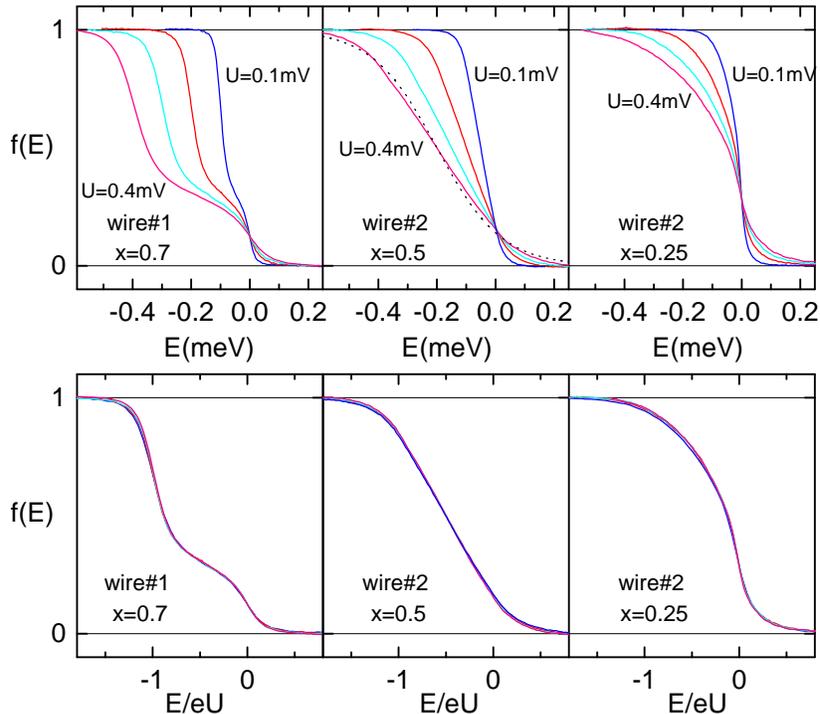, width=4.2419in, clip=}
\caption{Measured distribution
functions $f(E)$ on wire \#1 (left panel) and wire \#2 (central and right
panel) for $U=0.1,~0.2,~0.3$ and $0.4~\mathrm{mV,}$ plotted as a function of 
$E$ (top) and $E/eU$ (bottom)$\mathrm{.}$ The dotted line in the central
panel is the prediction in the hot electron regime for $U=$ $0.4~\mathrm{mV.}$}
\label{f(E)}
\end{figure}

We show in Fig.~\ref{f(E)} the distribution functions measured with the
three junctions at $U=0.1,$ $0.2,$ $0.3$ and $0.4$~$\mathrm{mV}$ \cite%
{heatingS}.

As expected from the
difference in the diffusion times, the distribution functions are much more
rounded in wire \#2 than in wire \#1.\ The distribution function at the
middle of wire \#2 is close to the hot electron prediction, shown with a
dotted line in the central panel for $U=0.4~\mathrm{mV.}$ However, the
distribution functions measured on the same wire close to the left electrode
(right panel) are clearly different from Fermi functions: they display a
strong slope near $E=0,$ reminiscent of the Fermi step at this energy in the
closest pad (A) which was grounded.

The larger the voltage $U,$ the wider the interval of energy in which $f$
varies from 1 to 0, as expected in both limiting regimes (see Fig.~\ref%
{FigSetup}). In order to remove this dependence, we have replotted the same
data in the bottom of Fig.~\ref{f(E)}, but with the reduced energy $E/eU$ on
the horizontal axis. Remarkably, all scaled curves superimpose. This scaling
property, which was also found in all the experiments on copper wires, is
not generic: in experiments on silver wires, the slope of $f(E)$ at $%
E/eU=-0.5$ was found to increase with $U$ \cite{relaxAg}$.$

\subsection{Energy exchange kernel}

The distribution function can be calculated by solving the stationary
Boltzmann equation in the diffusive regime \cite{Nagaev,Kozub}: 
\begin{equation}
\frac{1}{\tau _{D}}\frac{\partial ^{2}f\left( x,E\right) }{\partial x^{2}}+%
{\mathcal{I}}_{\mathrm{coll}}^{\mathrm{in}}\left( x,E,\left\{ f\right\}
\right) -{\mathcal{I}}_{\mathrm{coll}}^{\mathrm{out}}\left( x,E,\left\{
f\right\} \right) =0  \label{Boltzmann}
%f\right\} \right) =0  \tag{2}  \label{Boltzmann}
\end{equation}%
where ${\mathcal{I}}_{\mathrm{coll}}^{\mathrm{in}}\left( x,E,\left\{ f\right\}
\right) $ and ${\mathcal{I}}_{\mathrm{coll}}^{\mathrm{out}}\left( x,E,\left\{
f\right\} \right) $ are the rates at which quasiparticles are scattered in
and out of a state at energy $E$ by inelastic processes. The observation of
the hot-electron regime in the middle of wire \#2, with a temperature close
to the expected one, indicates that the energy is mainly redistributed among
the quasiparticles. In particular, phonon emission can be neglected.
Assuming that the dominant inelastic process is a two-quasiparticle
interaction which is local on the scale of variations of the distribution
function, 
\begin{equation}
{\mathcal{I}}_{\mathrm{coll}}^{\mathrm{in}}\left( x,E,\left\{ f\right\}
\right) =\int {\mathrm{d}}\varepsilon {\mathrm{d}}E^{\prime }K\left( \varepsilon
\right) f_{E+\varepsilon }^{x}(1-f_{E}^{x})f_{E^{\prime
}}^{x}(1-f_{E^{^{\prime }}-\varepsilon }^{x})  \label{Iout}
%}}^{x}(1-f_{E^{^{\prime }}-\varepsilon }^{x})  \tag{3}  \label{Iout}
\end{equation}%
where the shorthand $f_{E}^{x}$ stands for $f\left( x,E\right) .$ The
out-collision term ${\mathcal{I}}_{\mathrm{coll}}^{\mathrm{out}}$ has a
similar form. The kernel function $K\left( \varepsilon \right) $ is
proportional to the averaged squared interaction between two quasiparticles
exchanging an energy $\varepsilon .$ We have neglected the possible
dependence of $K(\varepsilon )$ on the energies of the initial and final
states and on the position along the wire. The theory of electron-electron
interactions in diffusive conductors in the $1\mathrm{D}$ regime \cite{AA}
predicts $K\left( \varepsilon \right) \propto \varepsilon ^{-3/2},$ a regime
which was observed in silver wires \cite{relaxAg}. In gold and copper wires,
the scaling property implies, by a simple change of variables in Eq.~(3),
that $U^{2}K\left( \varepsilon \right) $ is a function of $\varepsilon /eU$
only \cite{Relax_1}.\ If futhermore $K\left( \varepsilon \right) $ does not
depend on $U,$ one obtains $K\left( \varepsilon \right) =\gamma /\varepsilon
^{2},$ with $\gamma $ a typical interaction rate. We nevertheless tried to
fit our data at $U=0.1~\mathrm{mV}$ with $K\left( \varepsilon \right)
=\kappa /\varepsilon ^{3/2}$, and obtained $\kappa =50~\mathrm{ns}^{-1}%
\mathrm{meV}^{-1/2}$, which is three orders of magnitude larger than the
theoretical prediction $\kappa ^{\mathrm{thy}}=(\hbar ^{3/2}\nu S\pi \sqrt{%
D/2})^{-1}\approx 0.06~\mathrm{ns}^{-1}\mathrm{meV}^{-1/2},$ with $\nu $ the
density of states in gold and $S$ the cross-section of the wire \cite%
{Kamenev,relaxAg}. Moreover, the shape of the distribution functions on wire
\#2 is not properly reproduced and the calculated curve for wire \#1 at $%
U=0.4~\mathrm{mV}$ with the same parameter is significantly more rounded
than the experimental data.

\subsection{Fits}

\begin{figure}[!ftbh]
\centering
\epsfig{file=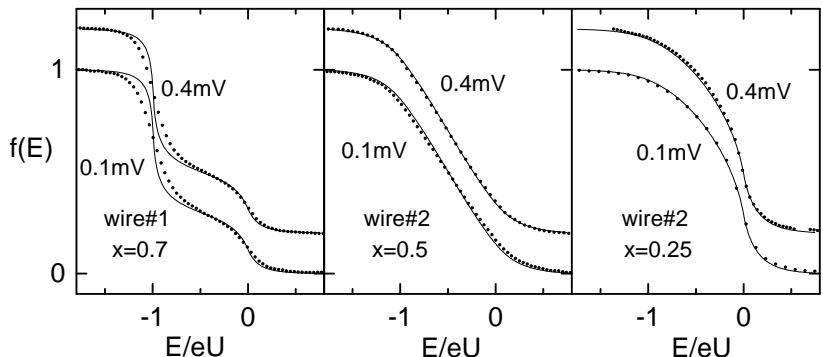, width=4.2748in, clip=}
\caption{Symbols:
measured distribution functions with $~U=0.1~\mathrm{mV}$ and (shifted
vertically) $U=0.4~\mathrm{mV.}$ Solid lines: calculated distribution
functions with the phenomenological kernel $K\left( \protect\varepsilon %
\right) =\protect\gamma /\protect\varepsilon ^{2}.$ Reservoirs are here
assumed to be at $T=33~\mathrm{mK.}$}
\label{fitsexp2}
\end{figure}

Figure \ref{fitsexp2} shows the best fit of the data with $K\left(
\varepsilon \right) =\gamma /\varepsilon ^{2},$ obtained with $\gamma =8.9~%
\mathrm{ns}^{-1}$.~

\begin{figure}[!ftbh]
\centering
\epsfig{file=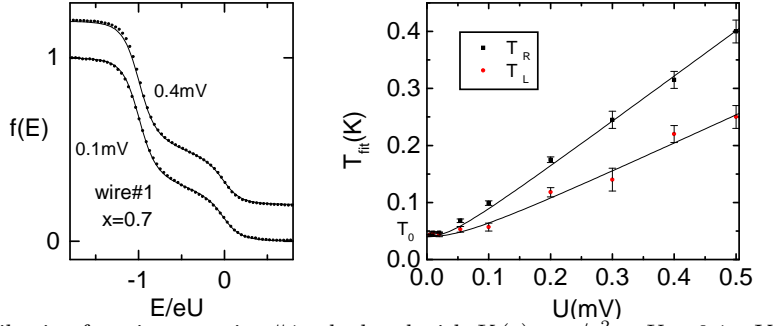, width=3.9211in, clip=}
\caption{Left
panel : distribution functions on wire \#1 calculated with $K\left( \protect%
\varepsilon \right) =\protect\gamma /\protect\varepsilon ^{2}$ at $U=0.1\ 
\mathrm{mV}$ and $0.4\ \mathrm{mV,}$ the reservoirs temperatures being taken
as fit parameters (solid lines). Symbols are data points. Right panel:
symbols: temperatures of the right ($T_{R})$ and left ($T_{L}$) reservoirs
obtained from the fits$.$ Solid lines: reservoir heating model with $b$ as a
free parameter.}
\label{tempreservoirs}
\end{figure}

By construction, the
scaling property of the data is exactly reproduced. Whereas the shape of the
distribution functions on wire \#2 is well accounted for, the experimental
data on wire \#1 are more rounded than we calculate.\ The opposite
discrepancy was observed in experiments on $5$~$\mu \mathrm{m-}$long copper
wires \cite{Cuunpublished}. We attribute the strong rounding in the
distribution functions on wire \#1 to the large heating of the reservoirs
associated with the low resistance of wire \#1: at a given value of voltage $%
U,$ the heat power $P=U^{2}/R$ is the highest in the less-resistive wires.
The solution of the heat equation in the contacts, assumed to have the shape
of an angular sector, with angle $\theta $, and neglecting phonon emission,
gives the reservoir temperature at the end of the wire \cite{Henny}: $T=%
\sqrt{T_{0}^{2}+b^{2}U^{2}}.$ Here, $T_{0}$ is the temperature of the
quasiparticles at a large distance $r_{\max }$ from the contact to the wire,
and $b=\sqrt{\frac{1}{\theta \mathfrak{L}}\frac{R_{\Box }^{\mathrm{res}}}{R}%
\ln \frac{r_{\max }}{r_{\min }}},$ with $R_{\Box }^{\mathrm{res}}\approx
0.05~\Omega $ the sheet resistance of the reservoir, estimated from the
resistivity of the wires and the thickness ratio of the wires and of the
reservoirs, $r_{\max }\approx 1~\mathrm{mm}$ a typical equilibration length
between electrons and phonons \cite{Roukes}, and $r_{\min }$ the smallest
radius for the radial approximation to be valid: $r_{\min }\approx w.$ In
the experiment, the left contact (labelled A in Fig.~\ref{dessin}), opens
with an angle $\theta _{L}\approx 2.44\mathrm{~rad},$ whereas the two other
contacts (B and C in Fig.~\ref{dessin}) have a smaller angle: $\theta
_{R}\approx 0.96\mathrm{~rad}.$ We have fitted the distribution functions
separately from $U=0.005\ \mathrm{mV}$ to $U=0.5\ \mathrm{mV,}$ with the
temperatures of the reservoirs as fit parameters. The results are shown in
Fig.~\ref{tempreservoirs}. 

At voltages below $20\ 
\mathrm{\mu V,}$ the fit temperature is nearly constant and identical in
both reservoirs: $T_{0}=43~\mathrm{mK,}$ in reasonable agreement with the
temperature $T=33~\mathrm{mK}$ indicated by the thermometer on the mixing
chamber. At such low voltages, scaling is not obeyed since temperature
produces a significant rounding of the steps and $eU/k_{B}T$ varies with $U.$
From the fit of the data on Fig.~\ref{tempreservoirs}, we deduce $b_{L}=0.50~%
\mathrm{mK/\mu V}$ for the left contact, and $b_{R}=0.80~\mathrm{mK/\mu V}$
for the right contact. Theory gives $b_{R}/b_{L}=\sqrt{\theta _{L}/\theta
_{R}}=1.60,$ in excellent agreement with the ratio found from the fits. Our
estimated values of $b_{L}$ and $b_{R}$ are about a factor of 2 larger than
those obtained from Fig.~\ref{tempreservoirs}, possibly because the sheet
resistance of the reservoirs is less than our estimate based on the
electronic mean free path in the wire. At voltages larger than $50~\mu 
\mathrm{V}$, the temperature of the reservoirs is proportional to $U,$
leading to a scaling in the rounding of the steps. In contrast, we want to
stress that the slope of the distribution function near $E/eU=-0.5$ cannot
be explained by heating alone. The fits of the distribution functions in
this regime, taking into account both $K\left( \varepsilon \right) =\gamma
/\varepsilon ^{2}$ (with $\gamma =8.9~\mathrm{ns}^{-1}$) and heating is
shown on the left panel of Fig.~\ref{tempreservoirs}. A consequence of
reservoir heating is that the distribution functions are rounded at the
scale of $\frac{k_{B}}{e}bU\approx U/15,$ and the sharp features that were
observed on copper samples cannot not be resolved \cite{Cuunpublished}.

\subsection{Two-level systems}

The distribution functions can also be well accounted for by a model in
which quasiparticles are in local equilibrium with two-level systems
distributed uniformly along the wire. The relevance of TLS on phase
relaxation has recently been suggested by several authors \cite%
{Imry,Zawa,Kroha}. Agreement is found if one assumes, in a simple model
(described in \cite{relaxAg}), that the quasiparticles are weakly coupled to
the TLS with a density inversely proportional to the spacing $\varepsilon $
between the two levels. Such a density is obtained if the two-level systems
are the two lowest energy levels in symmetric double-well potentials, and if
the distribution of the barrier heights is uniform. In glasses the
distribution of potential well asymmetries is usually taken as white~\cite%
{glasses}, but one might argue that in metals, symmetric double-wells could
result from crystalline symmetries~ \cite{delft}. We note that a similar
assumption is made in calculations based on a strong coupling to TLS \cite%
{Kroha}.

\section{Resistance measurements}

Complementary information on interactions was obtained from resistance
measurements. We have fabricated long gold wires in the same deposition
machine as for the energy relaxation experiment. To enhance adhesion of the
Au film on the substrate, we used two different methods. For a first sample
(Au1)~ \cite{WLGoteborg}, we evaporated first $1~\mathrm{nm}$ of aluminum,
and oxidized it. On a second sample (Au2), the surface of the sample was
ion-milled just before gold deposition. A more complete set of data was
taken on sample Au2, and we report only here the results on this sample. Its
length, width and thickness are $L=271~\mu \mathrm{m,}$ $w=115~\mathrm{nm,}$
and $t=45~\mathrm{nm,}$ respectively. The low temperature resistance was $%
1125~\Omega .$ Assuming Einstein's relation and a rectangular cross-section,
we deduce the diffusion constant $D=0.016$~\textrm{m}$^{2}$\textrm{/s }\cite%
{RRR}\textrm{. }In another fabrication run conducted at Michigan State
University, we have Joule evaporated another sample, called AuMSU, with very
pure gold (99.9999\%, \emph{i.e. }$1~\mathrm{ppm}$ of impurities), with $%
L=176~\mu \mathrm{m,}$ $w=80~\mathrm{nm}$, $t=45~\mathrm{nm}$ and $D=0.016$~%
\textrm{m}$^{2}$\textrm{/s}. All the samples were measured in the same
top-loading dilution refrigerator.

\subsection{Resistance vs. temperature}

\begin{figure}[!ftbh]
\centering
\epsfig{file=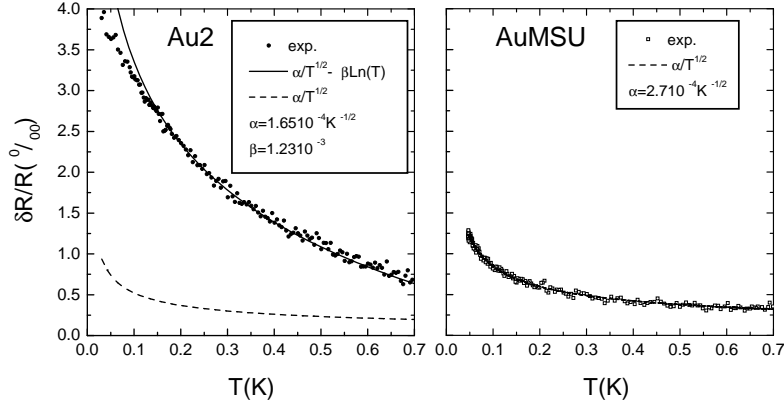, width=4.094in, clip=}
\caption{Relative variations of the resistance of sample
Au2 and AuMSU as a function of temperature. The dotted line is the
prediction of the theory of electron-electron interactions; the solid line
on the left panel is a fit with an additional logarithmic term.}
\label{RvsT}
\end{figure}

We show in Fig.~\ref{RvsT} the temperature dependence of the resistance for
samples Au2 and AuMSU. The contribution of weak localization for Au2,
smaller than $10^{-4}$ (see below), can be neglected. For AuMSU, this
contribution, up to 3$\times 10^{-3}$ in $\delta R/R,$ has a well understood
temperature dependence (see below), and was substracted. The contribution of
electron-electron interactions to the resistivity, $\frac{\delta R_{ee}}{R}%
\approx 3.13\frac{R}{R_{K}}\frac{L_{T}}{L}=\frac{\alpha }{\sqrt{T}}$ \cite%
{Waves}, with $R_{K}=\frac{h}{e^{2}}$ and $L_{T}=\sqrt{\frac{\hbar D}{k_{B}T}%
},$ plotted as dotted lines in Fig.~\ref{RvsT}, accounts well for the data
of AuMSU \cite{ReeOKinAgCu}, where the fit parameter $\alpha =2.7\times
10^{-3}~K^{-1/2}$ is very close to the calculated value $\alpha ^{\mathrm{thy%
}}=2.5\times 10^{-3}~K^{-1/2}.$ In contrast, the variations observed in Au2
are stronger and have a different temperature dependence.~

A good fit of the data can
be obtained with: $\frac{\delta R}{R}=\frac{\delta R_{ee}}{R}-\beta \log T,$
with $\beta =1.23\times 10^{-3}.$ The logarithmic term is characteristic of
the Kondo effect \cite{Kondo}, and it was in particular studied in great
detail in dilute alloys of Fe in Au \cite%
{Giordano,Bergmann,Chandrasekhar,MWKondo}, where $\beta $ was found to be
proportional to the Fe concentration and dependent on the cross section of
the wire as well as on disorder. Assuming that the logarithmic dependence of
the resistance of Au2 is due to Fe impurities, we get, by comparison with a
sample with similar width, thickness and diffusion constant (AuFe2 in \cite%
{MWKondo}), an impurity concentration $c\approx 55~\mathrm{ppm,}$ which is
compatible with the purity of the gold source used for evaporation (99.99\%, 
\emph{i.e. }$100~\mathrm{ppm}$ of impurities) \cite{SIMS}.

\subsection{Resistance vs. magnetic field - Phase coherence time}

\begin{figure}[!ftbh]
\centering
\epsfig{file=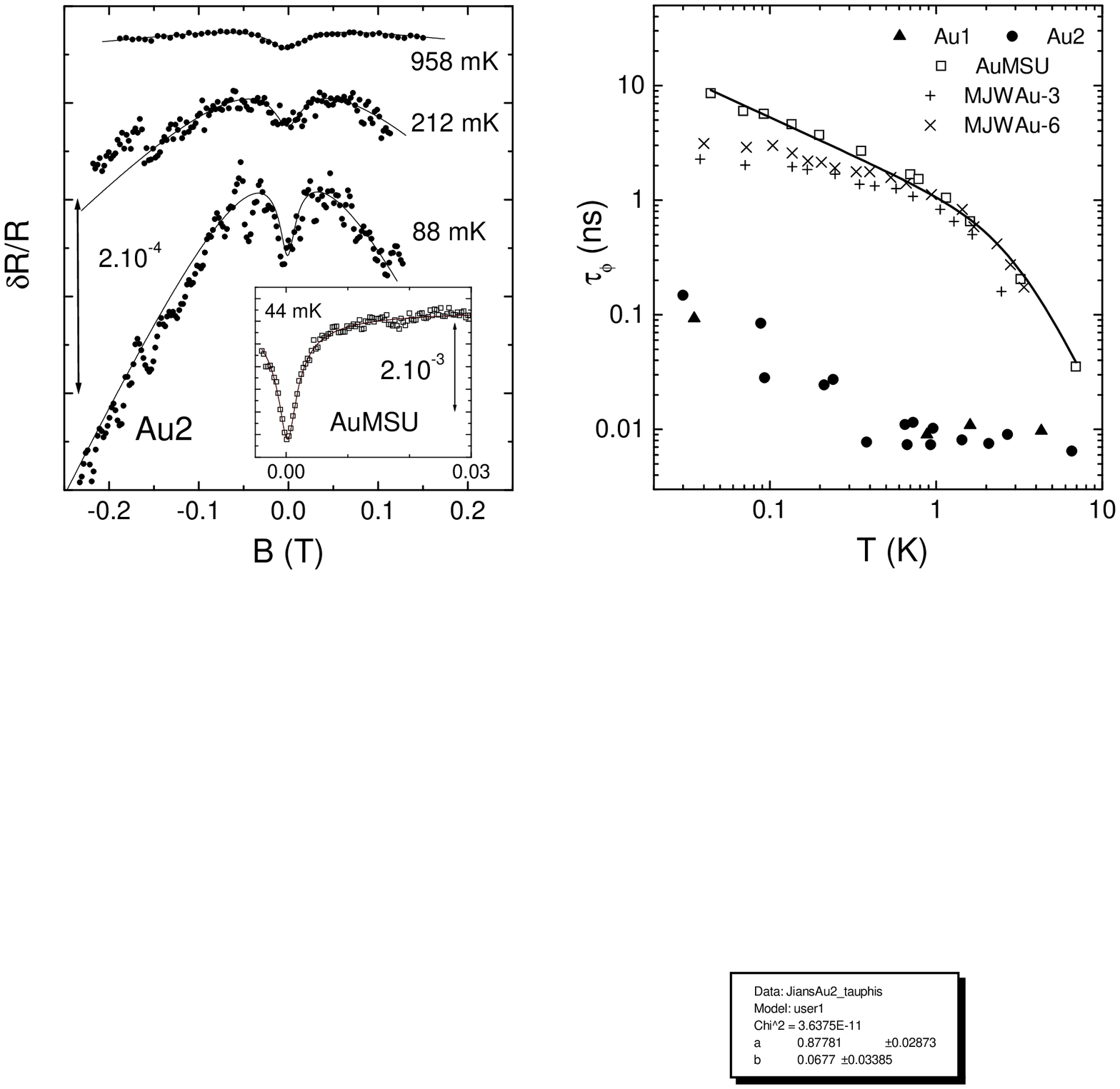, width=4.7202in, clip=}
\caption{Left panel: Symbols:
magnetoresistance of sample Au2 (vertical shifts between the curves taken at
different temperatures are arbitrary) and of sample AuMSU (inset; note the
difference in the scales). Solid curves: fit with the sum of the weak
localisation expressions and (for Au2 only) of a phenomenological large
scale contribution $-r(T)\protect\sqrt{B_{0}^{2}+B^{2}}.$ Right panel: phase
coherence time as a function of temperature, for samples Au1, Au2 and
AuMSU.\ Solid line: fit of AuMSU with $\protect\tau _{\protect\phi %
}^{-1}=AT^{2/3}+BT^{3}.$ For reference, we have also plotted the data of two
samples of Ref. \protect\cite{MW} similar to AuMSU (MJW Au-3\&6)}
\label{RvsB}
\end{figure}

We present in the left panel of Fig.~\ref{RvsB} the magnetoresistance of Au2
and AuMSU.~
Around zero magnetic field,
the resistance presents a dip, as expected from weak antilocalisation \cite%
{AA}. A large-scale negative magnetoresistance is found on Au2, which is
similar to measurements performed on dilute alloys of Fe in Au \cite%
{Giordano,MWKondo}. It can be attributed to the freezing of the magnetic
moment of the impurities with the magnetic field \cite{Giordano}. In order
to extract the phase coherence time from the magnetoresistance of Au2, we
subtracted out the large scale magnetoresistance, which was, at each
temperature, fitted with $\Delta R(B)/R=-r(T)(\sqrt{B_{0}^{2}+B^{2}}-B_{0}).$
The remaining magnetoresistance was then fitted with the predictions of the
weak localization theory \cite{AA,WLGoteborg}. In the right panel of Fig.~%
\ref{RvsB}, we show the phase coherence time $\tau _{\varphi }$ of samples
Au1, Au2 and AuMSU as a function of temperature. Upon cooling, the phase
coherence time of Au1 and Au2 remains unchanged from $8~\mathrm{K}$ to $0.5~%
\mathrm{K}$: $\tau _{\varphi }\approx 10~\mathrm{ps,}$ then it increases
roughly as $1/T$. A simple interpretation of the very low value of $\tau
_{\varphi }$ and of the desaturation below $0.5~\mathrm{K}$ is found by
comparing with similar measurements performed on gold wires in which a small
amount of iron was purposely introduced \cite{Giordano,Bergmann,MW,MWKondo}.
The same behavior of $\tau _{\varphi }(T)$ was found in these experiments,
and it was interpreted as an effect of spin-flip scattering by the
impurities, the rate of which presents a maximum at the Kondo temperature of
Fe in Au ($T_{K}\approx 0.3-1~\mathrm{K}$). The value $\tau _{\varphi }^{0}$
of $\tau _{\varphi }$ near the Kondo temperature was found to be roughly
given by $\tau _{\varphi }^{0}\approx 0.25~\mathrm{ns/}c,$ with the
concentration of impurities $c$ expressed in ppm (parts per million) \cite%
{MW,MWKondo}. In this interpretation, $\tau _{\varphi }^{0}\approx 10~%
\mathrm{ps}$ is obtained for $c\approx 25~\mathrm{ppm.}$ This provides us
with another estimation of $c$ which is of the same order of magnitude as
what we deduced from the resistance vs. temperature measurements. Sample
AuMSU does not show any magnetoresistance at large scale, and the phase
coherence time presents no saturation down to $44~\mathrm{mK}$ (see Fig.~\ref%
{RvsB}). A good fit of the data could be obtained with the theoretical
dependence \cite{AA} $\tau _{\phi }^{-1}=AT^{2/3}+BT^{3},$ with $A=0.9~%
\mathrm{ns}^{-1}\mathrm{K}^{-2/3}$ and $B=68~\mathrm{\mu s}^{-1}\mathrm{K}%
^{-3}.$ The theoretical value from the theory of electron-electron
interactions is $A=0.5~\mathrm{ns}^{-1}\mathrm{K}^{-2/3}$ \cite{Waves}$.$

To allow for a comparison with the results of Mohanty \emph{et al.}, we have
also plotted on the same figure the phase coherence time of samples Au-3 and
Au-6 in \cite{MW}, which have similar size and diffusion constants as AuMSU
(Au-3: $w=100~\mathrm{nm}$, $t=35~\mathrm{nm}$ and $D=0.011$~\textrm{m}$^{2}$%
\textrm{/s}; Au-6: $w=180~\mathrm{nm}$, $t=40~\mathrm{nm}$ and $D=0.016$~%
\textrm{m}$^{2}$\textrm{/s}). It is clearly seen that our data do not
display the saturation found in \cite{MW}. We had already found no
saturation in measurements on silver wires \cite{WLGoteborg}, which were
made from very pure (99.9999\%) Ag. We suspect that energy exchanges in
samples fabricated with very pure Au would be consistent with the theory of
Ref.~\cite{AA}, as in very pure Ag \cite{relaxAg}, but this experiment
remains to be performed. It is noteworthy that the low-temperature
saturation of $\tau _{\phi }$ reported in Cu wires \cite{WLGoteborg} was
identically found in other samples obtained from 99.9999\% Cu, supporting
the hypothesis that surface oxide could play a major role in Cu \cite%
{Haesendonck}.

\section{Can magnetic impurities mediate electron-electron interactions?}

The data presented in section 2 provide strong evidence that some of the
gold wires we have studied contain several tens of ppm of iron. The question
arises as to whether the energy exchange measurements can also be explained
by the presence of these magnetic impurities. Kaminski and Glazman \cite%
{Glazman} recently pointed out that $K(\varepsilon )\propto \varepsilon
^{-2} $ is obtained in second order perturbation theory from the interaction
of two quasiparticles with a magnetic impurity. The effective matrix element 
$M_{eff}$ of the second order process is proportional to $%
J^{2}/(E_{v}-E_{i}), $ where $J$ is the coupling parameter between the
quasiparticles and the magnetic impurity and $E_{v}-E_{i}$ is the energy
difference between the intermediate (virtual) state and the initial state.
In the intermediate state, one quasiparticle is promoted to an energy state
higher by $\varepsilon $ than the initial state, whereas the magnetic
impurity has been reversed. One thus obtains $K(\varepsilon )\propto c\times
\left| M_{eff}\right| ^{2}=cJ^{4}/\varepsilon ^{2},$ which has the energy
dependence observed in the experiment, but the bare coupling parameter $%
J\approx 1/\nu \log (E_{F}/k_{B}T_{K})$ is too small$.$ If one assumes that
we have Fe impurities in Au, $k_{B}T_{K}\approx 50~\mu \mathrm{eV}$ is of
the order of the energies probed in the energy exchange experiment, and the
coupling constant is expected to be renormalised by the Kondo effect. This
effect has been treated in \cite{Glazman}, and the proposed expression $%
K(\varepsilon )=\frac{\pi }{2}\frac{c}{\hbar \nu }S(S+1)\log ^{-4}(\frac{eU}{%
T_{K}})\frac{1}{\varepsilon ^{2}},$ has now the right order of magnitude but
too strong a $U$-dependence. Further work is clearly needed to clarify this
issue.


\begin{thebibliography}{99}
\bibitem{field} We generally find that the current voltage characteristic of
the junction presents a small hysteresis at voltages $\pm \Delta /e$, that
we attribute to surface effects on the probe electrode. When applying a
small magnetic field ($B\approx 5~\mathrm{mT}$) perpendicularly to the
surface of the chip, this singularity disappears. The magnetic field is kept
at the same value during all the experiment.
\bibitem{RRR} A slightly higher value $D=0.0185$~\textrm{m}$^{2}$\textrm{/s }%
is obtained from the resistivity ratio $R(300~K)/R(4.2~K)=$ $2.18.$
\bibitem{ReeOKinAgCu} The same agreement was obtained in measurements of Cu
and Ag wires.
\bibitem{SIMS} A Secondary Ions Mass Spectroscopy analysis on a sample
fabricated similarly to Au1 and Au2 concluded that the concentration of iron
was smaller than 100~ppm.
\end{thebibliography}
\end{document}